\documentclass[prl,aps,amssymb,showpacs,twocolumn]{revtex4}
\usepackage{amsmath}
\usepackage{amssymb}
\usepackage{amsthm}
\usepackage{amsfonts}
\usepackage{enumerate}
\usepackage{latexsym}
\usepackage{graphicx}

\newcommand{\ba}{\begin{eqnarray}}
\newcommand{\ea}{\end{eqnarray}}
\newcommand{\be}{\begin{equation}}
\newcommand{\ee}{\end{equation}}
\newcommand{\n}{\nonumber \\ }

\begin{document}

\tolerance 10000

\newcommand{\vk}{{\bf k}}

\title{Scenario for Fractional Quantum Hall Effect in Bulk Isotropic Materials}

\author{F. J. Burnell$^1$,  B. Andrei Bernevig$^{1,2}$ and D. P. Arovas$^{3}$ }

\affiliation{$^1$Department of Physics, Princeton University, Princeton, NJ 08544}

\affiliation{$^2$Princeton Center for Theoretical Physics, Princeton, NJ 08544}

\affiliation{$^3$Physics Department 0319, University of California at San Diego, La Jolla, CA 92093}

\begin{abstract}
We investigate the possibility of a strongly correlated Fractional
Quantum Hall (FQH) state in bulk three dimensional isotropic (not
layered) materials. We find that a FQH state can exist at low
densities only if it is accompanied by a staging transition in which
the electrons re-organize themselves in layers, perpendicular to the
magnetic field, at distances of order the magnetic length apart.
The Hartree energy associated to the staging transition is
off-set by the correlation Fock energy of the 3D FQH state. We
obtain the phase diagram of bulk electrons in a magnetic field
subject to Coulomb interactions as a function of carrier density and
lattice constant. At very low densities, the 3D FQH state exhibits a
transition to a 3D Wigner crystal state stabilized by phonon correlations.
\end{abstract}

\date{\today}


\maketitle

The Quantum Hall effect is intimately linked to the
low-dimensionality of the sample and the fact that the magnetic
field quenches the kinetic energy of the electron liquid in $2$
dimensions ($2$D). The  Quantum Hall effect has also been observed
in $3$ dimensions ($3$D) in the Bechgaard salts
\cite{Balicas1995,McKernan1995} but only at integer filling
factors and only in layered samples. The $3$D Quantum Hall states
can be explained as a series of $2$D states, weakly coupled so
that the band-width in the $z$-direction be smaller than the $2$D
Landau gap. So far, $3$D FQH states have not been observed, even
in layered samples, due to their low electron mobility and small
many-body gap.

Recent experimental interest in graphene, graphite and Bismuth has
focused on the Dirac nature of carriers and on the Integer Quantum
Hall (IQH) effect. The Dirac dispersion in these materials gives
electrons large mean-free paths and large Landau gaps. As a
result, large IQH plateaus are observed, even at room temperature.
A $3$D IQH effect has been predicted in graphite
\cite{Bernevig2007}. Band structure considerations suggest that,
at sufficiently strong magnetic field, the $n=0$ and $n=1$ Landau
levels of graphite are gapped and a single $3$D IQHE plateau
exists. The layered structure of graphite is essential to the
existence of the $3$D IQH. Recent experiments in Bismuth suggest
the possibility of a $3$D bulk strongly correlated electron state
in the fractional filling regime \cite{Behnia2007}. The experiment
reports quasi-plateaus in $\rho_{xy}$ at fields which are integer
times larger than the field required to drive the system in the
quantum limit.

These results are unexpected since Bismuth, unlike Graphite, is an
isotropic material. There are only two known scenarios for the
existence of a Quantum Hall effect in 3D: it can be either a
band-structure effect, such as in graphite, or an interaction
effect, where a $2k_F$ instability opens a gap at the one
dimensional Fermi level in the direction of the magnetic field.
The Fermi level is then pinned in the many-body gap and the system
exhibits a Quantum Hall effect. However, by construction, both
these scenarios lead to an integer, and not fractional state.
While it is easy to generalize the Laughlin \cite{laughlin1983}
and Halperin \cite{Halperin1983} states to a 3D FQH state
\cite{macdonald,Naud2001} with correlations between layers, they
were never found to be the variational ground-states in an
isotropic material. Contrarily, it was found \cite{macdonald} that
the magneto-plasmon gap closes and a phase-transition to a crystal
state occurs whenever the distance between the layers is roughly
smaller than half the magnetic length.

 Previous work on quantum well bi-layer
and multi-layer systems suggests several possible electronic
arrangements for a 3D material in a magnetic field: MacDonald
and coauthors \cite{JoyntMacD} considered the energetics of $3$D Halperin states, which are adaptations of the
Laughlin wave functions to ensure that electrons in adjacent
layers maximize their separations. At very small inter-layer
separations, they found these states crystalize due to a collapse
of the magnetoroton gap. MacDonald \cite{macdonald} also
established that in multilayers, in the
Hartree Fock approximation, the gain in exchange energy from
distributing electrons unequally between the layers can exceed the
electrostatic cost of increased inter-layer Coulomb energy leading
to a {\it staging transition}. Ref. \cite{MacDonald02} proposed a third
type of candidate state, the spontaneous inter-layer coherent miniband state,
consisting of a combination of lowest Landau level (LLL) states in
each layer which forms a band in $k_z$. However, no state
exhibiting a 3D FQH effect has been found as the ground-state of
an isotropic material in $\vec{B}$ field.

In this paper, we revisit these issues and try to address the
question  of whether a FQH effect could exist in the $3$D
compound. We propose a new variational ground-state, a correlated
$3$D FQH-staged state in an isotropic material. We compare four
different potential ground states for materials in $\vec{B}$ field
in the ultra-quantum limit with strong Coulomb interactions and
find that 3D FQH ground-states do not occur in isotropic materials
unless they are accompanied, at very low densities, by staging
transitions similar to the ones observed in graphite intercalation
compounds \cite{Dresselhaus1981} and predicted to occur in
multi-quantum-well systems by MacDonald \cite{macdonald}.  In the
parameter range relevant to both graphite and Bismuth, we obtain
the phase diagram as a function of the electron filling fraction,
the ratio of inter-layer separation to magnetic length, and the
magnitude of the $c$-axis hopping. At high carrier density and
zero c-axis hopping, a (staged) integer quantum Hall liquid is
energetically favorable.  As the filling in occupied layers
decreases below $\nu = 1$, liquid states with correlations between
layers can develop, and become energetically favored because of
their decreased inter-planar Coulomb energy. At extremely low
densities, staged crystal states tend to be energetically favored.
A SILC miniband state becomes favored over the staged integer
liquids, but not their fractional counterparts, at the $c$-axis
hopping relevant to graphite. Thus we find a parameter regime in
which fractionally filled, staged quantum Hall liquid states are
the expected ground states of the layered system.

Let us consider an isotropic material subject to a magnetic field
parallel to one of its crystallographic axes. The magnetic field
quenches the kinetic energy of electrons in lattice planes
(layers) perpendicular to the magnetic field. A plausible
scenario, borrowed from MacDonald's work on staged states in
multiple quantum wells \cite{macdonald} is that the relatively
small inter-layer separations in isotropic materials favor staged
states which consist of $n$ layers of electron gas of density
$\overline{\sigma}_e -\delta\sigma$, and one layer of density
$\overline{\sigma}_e + n \delta \sigma$, where
$\overline{\sigma}_e$ is the initial electron density per layer,
and $\delta \sigma$ is the staging density.
 For crystalline states, staging increases the distance
between electrons which lie directly above one another, decreasing
the crystal's Coulomb energy at small inter-layer separations; for
liquid states staging becomes advantageous when the gain in Fock
energy by increasing the electron density in the filled layers
outweighs the Hartree cost of distributing charge unevenly between
the layers.  Thus as the separation between layers decreases,
staged states become increasingly energetically favorable.

We calculate the staging Hartree energy as described in
\cite{macdonald}.  We model the system as a stack of planes, each
of uniform charge density $\sigma_i$, where $ \sigma_i = - n
\delta \sigma$ if $i | (n+1) $ and $ \delta \sigma $ otherwise.
This charge density is comprised of an immobile positive
background (ionic) charge density $\overline{\sigma}_e$ in each plane, and
a staged electronic charge density $-\overline{\sigma}_e+
\sigma_i$.  $n$ is called the {\it staging number}. The 3-d charge
density is conveniently described in Fourier space as a sum of Kroeneker
$\delta$ terms
 \ba \label{rho}
 \rho(q_z) &=& \rho_1(q_z)+ \rho_2(q_z) \n
 \rho_1(q_z) &=& \delta \sigma \frac{V}{d} \sum_j \delta(q_z, 2 \pi j /d) \n
 \rho_2 (q_z) &=& -\delta \sigma \frac{V}{d} \sum_j \delta(q_z, 2 \pi j/(n+1)d )
 \ea
 \noindent where $V$ is the total sample volume,
and $d$ is the lattice constant in the direction of the
magnetic field.  This charge density configuration has a Hartree
energy $E_{s}=\frac{e^2}{ \epsilon V} \sum_{q_z}
\frac{2\pi}{q_z^2}|\rho(q_z)|^2 $, or:
 \ba \label{En2}
 E_{s}&=&  \frac{ e^2}{\epsilon l}\frac{(\delta \nu)^2}{
  \nu_0}  \frac{d}{12 l} (n^2+2n)
   \ea \noindent
where $\nu_0 \equiv 2 \pi l^2 \overline{\sigma}_e$, and $\delta \nu = 2 \pi
 l^2 \delta \sigma $ is the electronic filling staged
from each layer.  One could also consider multiply staged states,
in which three or more different staging numbers $n$ exist.
However, we numerically found that the states of lowest energy are
those with charge distributions of the form in Eqn. (\ref{rho}),
described by a single staging number.

The first class of candidate states that we consider are the
staged liquid states.  The simplest such staged liquids consist of
$n$ layers depleted of electrons, and one layer of
electronic charge density $\overline{\sigma}_e+ n\delta\sigma$
(which we will call the occupied layer) in a quantum Hall liquid
state.  The total energy of a staged liquid
state is then:
 \begin{equation} \label{LiquidE}
   E = \frac{n}{n+1} E_{l}(\nu_0 - \delta \nu)+
\frac{1}{n+1} E_{l} (\nu_0 + n \delta \nu) +E_{s}
 \end{equation}
where $n$ is the staging number, $E_l (\nu)$ is the energy of the
liquid at filling $\nu$, $E_s$ is the Hartree energy
given by (\ref{En2}), $\nu_0$ is the mean filling factor, and $0
\leq \delta \nu \leq \nu_0$ is the amount of charge staged out of
each layer.

As Laughlin states have good in-plane correlation energies,
we consider first the case where the occupied layers are in a
Laughlin state.  We will refer to
Laughlin states at filling fraction $1/m$ in the occupied layers
as $(0,m,0)$ states.  Since the filling fraction in these layers is then
fixed at a value of $\nu = 1/m$ for some odd integer $m$, the quantities $\nu,
\nu_0$ and $n$ are related by $n +1= \frac{\nu}{\nu_0}$. Using
(\ref{LiquidE}), the optimal staged liquid state for a given
$\nu_0$ and $d/l$ can be found by optimizing over $\delta
\nu$ and  $n$. We find that optimal states in the regimes of
interest are fully staged ($\delta \nu = \nu_0$), so that the
first term of the right-hand side of (\ref{LiquidE}) vanishes.

\begin{figure}
\begin{center}
\includegraphics[width=3.4in, height=1.9in]{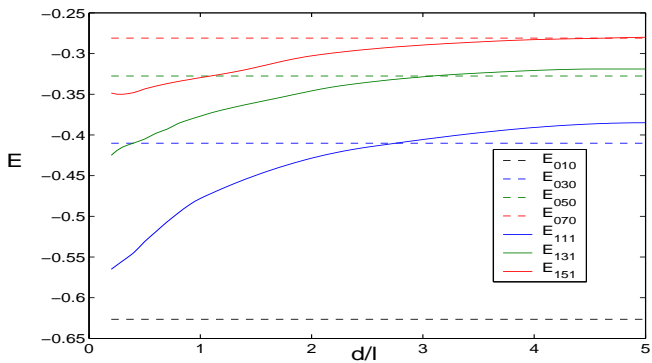}
 \end{center}
\caption{Energies of unstaged Laughlin (dashed lines) and Halperin
(solid lines) liquid states as a function of inter-layer
separation over magnetic length $(d /l)$. All energies are
reported as energies per electron, in units of $
\frac{e^2}{\epsilon l}$, where $l$ is the magnetic length. At
large separation the energy increases with $d/l$ as the
correlations in Eqn. (\ref{HalLiq})  do not optimally minimize the
intra-planar Coulomb interactions and their energy grows (see
text).  The minimum energy occurs when correlations effectively
separate electrons from both their in-plane and out-of-plane
neighbors.}\label{Tab1}
 \end{figure}

At small $d/l$, the simple uncorrelated liquid state
described above gives way to liquids with inter-layer correlations
\cite{JoyntMacD}, which we call Halperin liquid states. These
states have wave functions of the form
 \be \label{HalLiq}
\psi = \prod_{k;i, j} (z_i^{(k)} -z_j^{(k-1)} ) ^{m_1}(z_i^{(k)}
-z_j^{(k+1)} ) ^{m_1} \prod_{i < j} (z_i^{(k)} -z_j^{(k)} ) ^{m_2}
 \ee
where $k$ indexes the layer, and $i,j$ index particles within a
layer.  We refer to states of the form (\ref{HalLiq}) as $(m_1,
m_2, m_1)$ states.  Fig. [\ref{Tab1}] shows the energies for
several of these states as a function of inter-layer separation
and density, calculated using Monte Carlo techniques (see Appendix).

When $d/l$ is very small, one can imagine a low energy ground state of
staged Halperin liquid states, with $n$ layers completely depleted
of electronic charge for every occupied layer at filling $\nu =
1/(2 m_1+m_2)$.  In the layered system with a given mean filling
per layer $\nu_0$, the staging number is fixed by $n+1 =
\frac{1}{\nu_0(2m_1+m_2)}$. This in turn fixes the separation $(n+1)
d$ between occupied layers, and the energy of such a state
is
 \be
E = E_{l} ( (n+1) d) + E_{s} (d, n, \nu)
 \ee
where $E_{l} ((n+1) d)$ is the Coulomb energy of the state
(\ref{HalLiq}) at inter-layer separation $(n+1) d$, extrapolated
from the Monte Carlo results of fig.~\ref{Tab1}, and $E_{s}$ is
the Hartree staging term.   Fig.~\ref{Tab1} shows an important
property of the energies as a function of $d/l$. The wave function
(\ref{HalLiq}) contains correlations which separate electrons in
adjacent layers, at the expense of decreasing their average
separation in a given layer. Since these correlations do not
depend on $d/l$ (the out of plane correlations in Eqn.
(\ref{HalLiq}) depend on $z_i^{(k)} - z_j^{(k-1)}$, i.e. only on
the in-plane distance between particles in different layers), as
$d/l$ increases the energy of the Halperin states also tends to
increase. The inter-planar Coulomb interaction becomes weaker, so
that the energetic gain from inter-layer correlations decreases,
while the energetic cost of decreased separation in-plane relative
to the Laughlin state of the same filling remains constant. At
large $d/l$ the Laughlin liquid becomes the ground state.

In two dimensions, it is well known that for a low density of
carriers, a transition occurs from a FQH state to a triangular lattice Wigner
crystal of lattice constant $a =  l \left(\frac{\sqrt{3}\pi}{\nu}\right)^{1/2}$.
The Wigner crystal is energetically favored
over the FQH liquid state at fillings $\nu < 1/7$ \cite{LamGirvin}.  In multi-layer
systems at small $d$ ($d / l < 0.9 \sqrt{\frac{2
\pi}{\nu}}$), inter-layer repulsions favor an in-plane lattice
that is square as this permits larger separations between sites in
neighboring planes \cite{JoyntMacD}. As $d$ decreases,
multiple phase transitions between different stackings of squares
in the vertical direction occur; reference \cite{JoyntMacD}
identifies states in which electrons at the same in-plane
co-ordinates are separated by $M=2$, $4$ or $5$ layers, as well as states
with incommensurate stackings.  In this work we consider only true
crystal states with well-defined spatial periodicity, and will not include structures with
incommensurate stackings.

Numerical studies \cite{LamGirvin} indicate that the total energy
of these crystalline states is well approximated by
 \be \label{Ecrys}
E = E_0\, \nu^{1/2} + E_2\, \nu^{3/2} + E_3\, \nu^{5/2}.
 \ee
Here $E_0$ is the classical contribution, given by the energy of
the Madelung sum, $E_2$ is the phonon contribution, and $E_3$ is
an extra higher moment contribution calculated by a fit to a numerical evaluation of the Coulomb
energy.  For the low fillings at which Wigner crystals are
expected to occur, the total energy is well approximated by the
first two terms.

The classical contribution $E_0$ can be calculated using the Ewald
method, described in detail in \cite{deLeeuw}. We slightly modify
this method to account for the fact that the background charge is
localized in the planes. At each value of $d$, energies were
tabulated for a variety of crystal stackings and stagings; the
configurations of lowest energy were selected. We estimate $E_3$
by numerically extrapolating the results of \cite{JoyntMacD} as a
function of $n d/l$ to the values of interest here ($(n+1) d/l
\approx 0.2$ for most optimally staged crystal states).

We compute the contribution of lattice vibrations $E_2$ by
postulating a Lam-Girvin \cite{LamGirvin} variational wave function
of the form
 \be \label{psi}
\Psi= \exp\left ( \frac{1}{4} \sum_{ij} \xi_i\, B_{ij}\, \xi_j
\right ) \prod_j \phi_{{\vec R}_j}({\vec r_j})
 \ee
Here ${\vec R_j} = (X_j, Y_j)$ is the position of the lattice
site, $\xi_j \equiv (x_j-X_j)+i(y_j-Y_j)$ is the deviation of the
particle's position from this site in two-dimensional complexified
coordinates, and $\phi_{\vec R}({\vec r})$ are the lowest Landau
level coherent states,
 \be
 \phi_{\vec R}\big({\vec r}\big)=\frac{1}{\sqrt{ 2\pi l^2 }}\, e^{-({\vec r}-{\vec R})^2/4 l^2}
 e^{i{\hat z}\cdot{\vec r}\times{\vec R}/2l^2}
 \ee
The correlation matrix $B$ is a variational parameter; minimizing
the variational energy gives
 \ba \label{phonon}
B_{\vec k} =\frac{\omega_L ({\vec k}) - \omega_T ({\vec
k})}{\omega_L ({\vec k})  \, \omega_T ({\vec k})} e^{i \theta
_{\vec k}} \n E_2= \frac{m^*_e l^2}{4} \sum_{\vec k}
\left(\omega_L({\vec k}) + \omega_T({\vec k})\right)^2
 \ea
where $\omega_L$ and $\omega_T$ are the longitudinal and
transverse phonon frequencies, respectively, and $E$ is the energy
of the phonon modes.  The phonon frequencies are calculated using
the method of \cite{Clark}. For multiple-site unit cells, we fix
$B_{k}$ as a single function independent of site indices within
the unit cell.  In this case the energy is given by
(\ref{phonon}), with $\omega_T, \omega_L$ replaced by the square
roots of the eigenvalues of the $2 \times 2$ matrix formed by
averaging the second order correction to the Coulomb potential
over all sites in the unit cell.

\begin{figure}[!t]
\begin{center}
\includegraphics[width=3.4in, height=1.9in]{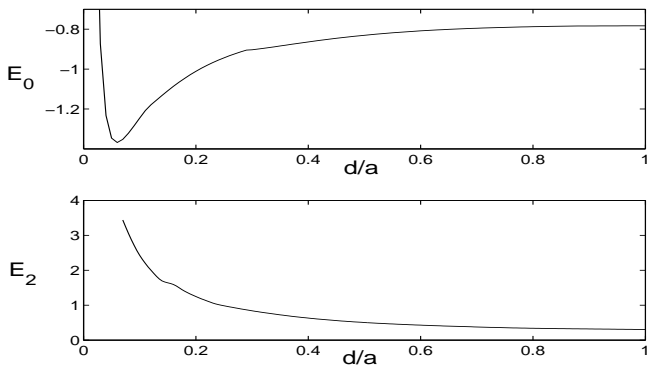}
\end{center}
 \caption{Calculated energies (per electron) of the unstaged crystalline states,
 as a function of $d/a$ where $a$ is the crystal lattice
 constant.  Energies are shown in units of $ \frac{e^2}{\epsilon l}$.  Top:
Coulomb sum contribution ($E_0$).  As $d/a$ decreases several phase transitions occur
 to crystal stacking patterns with a larger unit cell in the $z$
 direction, consistent with the findings of \cite{JoyntMacD}.
Bottom:  Phonon contribution ($E_2$) for the crystal states of lowest $E_0$.  As
 $d/a$ decreases the mean inter-electron distance decreases
 and $E_2$ grows.}
 \label{TabCrys}
\end{figure}

Fig.~\ref{TabCrys} shows the energies $E_0$ and $E_2$ of the
crystalline states over a range of inter-layer separations. At
very small $d/a$, the dominant interaction for a $M$-layer
crystal stacking is between an electron and its translates $M$
layers above and below; the associated energy increases roughly as
$\frac{1}{M d}$.  To avoid this cost, the crystal undergoes
several transitions to stackings of higher $M$ over the range of
inter-layer separations shown in fig.~\ref{TabCrys}
\cite{JoyntMacD}; at sufficiently small $d/a$ even the $M=5$
stacking does not adequately separate charges from their closest
vertical neighbors and the energy grows rapidly.  Though we
computed the energy of all three crystal structures ($2$- $4$- and
$5$- layer crystal stackings) at each value of $d/a$, only
the lowest of these energies is shown in the fig.~\ref{TabCrys}.

At the very small values of $d/l$ found in graphite and
bismuth, none of these crystal structures represent a stable
ground state. We therefore allow for Wigner crystals with staged
charge densities, requiring that charge depleted layers be
completely emptied of electrons ($\delta \sigma =
\overline{\sigma}_e)$.  The total energy for the staged Wigner
crystal is given by
 \be \label{EcrySt}
E = E_{\rm crys} + E_{s}
 \ee
where $E_{\rm crys}$ is given by (\ref{Ecrys}), using the values of
$E_0$ and $E_2$ shown in fig.~\ref{TabCrys} at the effective
inter-layer separation $(n+1) d$, and $E_{s}$ is given by
(\ref{En2}) with $\delta \nu = \nu_0$.  At a given inter-layer
separation $d$ and mean filling $\nu_0$, we find the
energetically optimal Wigner crystal by choosing the staging
number which minimizes (\ref{EcrySt}).

\begin{figure}[!t]
\begin{center}
\includegraphics[width=3.4in, height=1.9in]{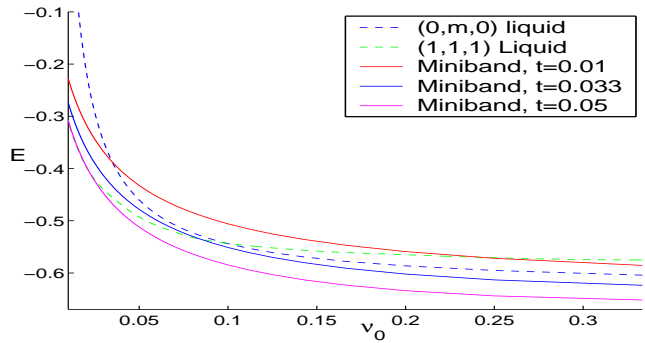}
\end{center}
\caption{ A comparison of the energies of the $(0,1,0)$ and $(1,1,1)$ staged
liquids with those of SILC miniband states as a function of mean
filling for $d/l=0.1$.  Energies and $t_\perp$ are shown in units of $ \frac{e^2}{\epsilon l}$.
At the physical value $t_\perp =0.033\cdot e^2/\epsilon l$ the SILC miniband state
is energetically favored at high densities, and has lower energy than the staged
integer-filled liquid state. The
$\nu_0$ dependence of the liquid energies can be understood from
fig.~\ref{Tab1} by noting that at fixed $d/l$ and
occupied filling $\nu$, $n d /l = \nu d/l\nu_0)$.  }
 \label{AllEComp}
\end{figure}

The final candidate state we consider is the Spontaneous Inter-Layer Coherent
State (SILC) miniband of Hanna, D{\'i}az-V{\'e}lez, and MacDonald \cite{MacDonald02}.  The SILC is given by the
Slater determinant state
 \be \label{psiSl}
 |\psi \rangle = \prod_{q, X} C^\dag_{q, X} |0\rangle
  \ee
where the creation operator generates the state
 \ba
\langle {\vec r} | C^\dag_{q, X} | 0 \rangle &= &\frac{1}{\sqrt{N_{\rm p}}} \sum_{j=1}^{N_{\rm p}}
e^{i jq d  } \psi_{j ,X}({\vec r})\\
\psi_{j,X} ({\vec r})  &=& \frac{1}{\sqrt{2 \pi
l^2}}\,\chi(z-jd)\, e^{iXy/l^2}\,e^{-(x-X)^2/2 l^2}\ ,\nonumber
 \ea
Here, $X$ is the center of the Landau strip, and $\chi(z-jd)$
describes the vertical localization of the electron to the $j^{\rm
th}$ plane; $N_{\rm p}$ is the number of planes (layers).  The
product $X$ is over all states in the lowest Landau level (LLL),
but only the lowest energy states in the momentum band are filled:
$- \pi \nu /d < q \leq \pi \nu /d$.

The energy of the SILC is given by \cite{JoyntMacD}
 \be
\frac{E_c}{N_e} = \frac{\nu e^2}{\epsilon l} \sum_j
\int \frac{d^2 r}{ 4\pi l^2} e^{-r^2/2 l^2} \left (\frac{\sin(\pi j \nu ) }{\pi j
\nu} \right )^2  -\frac{2 t_\perp \sin (\pi
\nu)}{\pi \nu}
 \ee
where $t_\perp$ is the hopping matrix element in the stacking direction.
Fig.~\ref{AllEComp} compares the energies of the SILC miniband states
at $d=0.1$ for several $t_\perp$ values with the energies of the
staged liquid states.  The energy of the SILC miniband state depends
strongly on the c-axis hopping matrix element $t_\perp$.  For
$t_\perp =0.01\frac{e^2}{\epsilon l}$, at the values of $d$
pertinent to graphite the SILC miniband state is never a ground state, as shown in
the fig.~\ref{AllEComp}. At intermediate values ( $t_\perp = 0.03\frac{e^2}{
\epsilon l}$), the SILC miniband state is the ground state
for sufficiently high densities. For $ t_\perp \geq 0.05
\frac{e^2}{\epsilon l}$,  the SILC miniband state is the ground
state for all densities shown.

We now compare the energies of the three types of states outlined
before in a parameter regime experimentally attainable for
graphite. We model graphite as a quantum Hall multilayer of
graphene sheets. Within each graphene plane, we assume that the
magnetic length is sufficiently large that the positive charge of
the graphene crystal is well approximated as a uniform surface
charge $\overline{\sigma}_e$. Further, we take the inter-plane
separation $d$ to be fixed, with a value of approximately
$d= 3.4\,$\AA.  We consider relatively strong magnetic
fields ($B \approx 30\,$T).

In graphite, at $B = 30\,$T, the relevant length scales is $l = 47\,$\AA,
giving $d/l \approx 0.07$.  The $c$-axis hopping in the LLL, as calculated from the $c$-axis bandwidth of the LLL \cite{Bernevig2007}, is
$t_\perp \approx 10\,$meV and the Landau gap is given by $E_{\rm gap}
= 30\,$meV.   The energy scale is set by $\frac{e^2}{
\epsilon_0 l} = 0.31\,$eV.  We assume that the mean filling fraction
can be tuned independently of the magnetic field.

\begin{figure}[!t]
\begin{center}
\includegraphics[width=3.4in, height=1.9in]{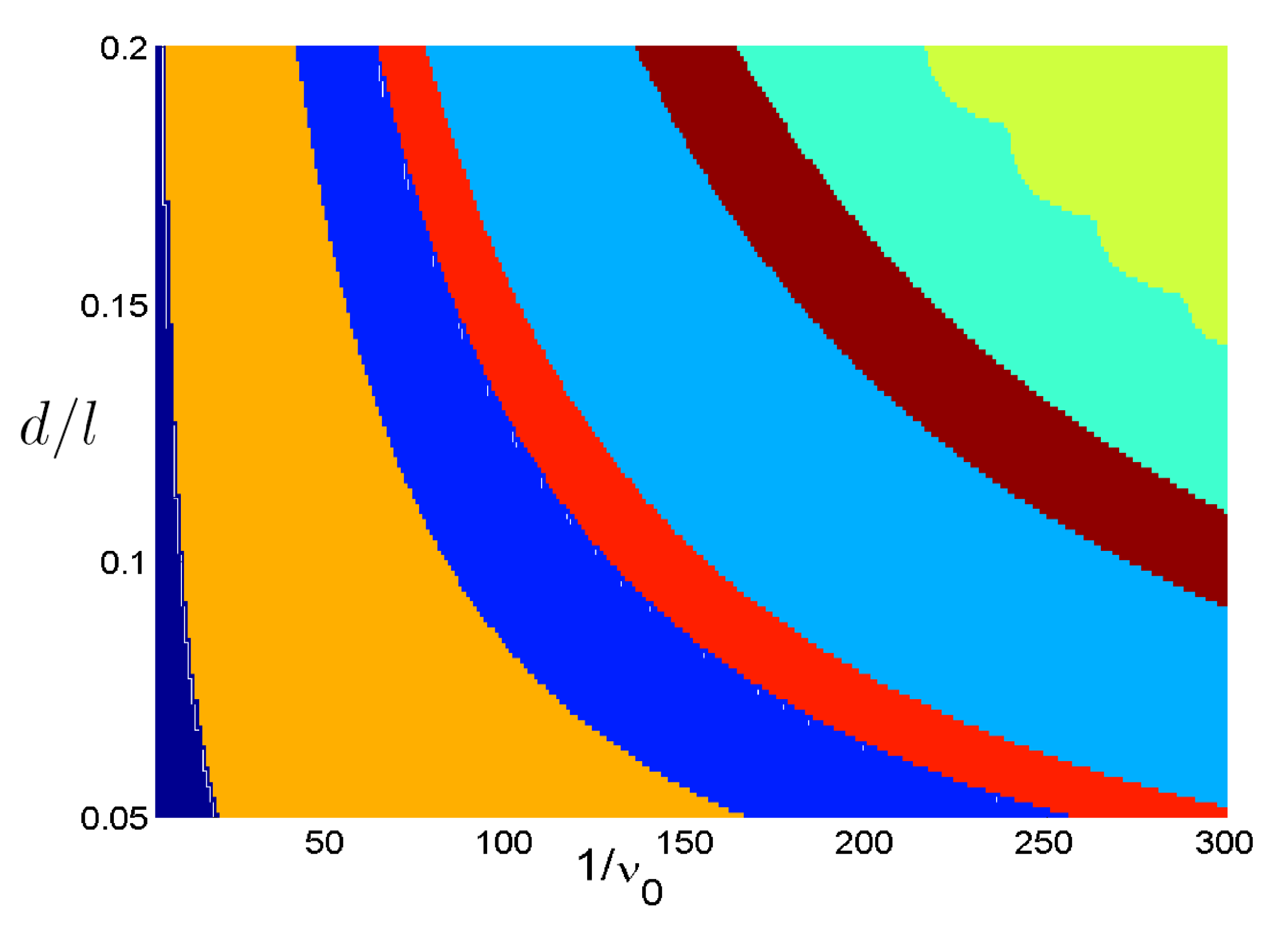}
\end{center}
\caption{Approximate phase portrait for graphite in the range $B
\approx 30 T$.  SILC miniband states are navy blue; the
Laughlin liquids are colored dark blue ((0,3,0) state), pale blue
((0,5,0) state), and turquoise ((0,7,0) state).  The Halperin
liquids are colored orange ((1,1,1) state), red ((1,3,1) state),
and brown ((1,5,1) state). Crystalline states are shown in green.
} \label{PhasePort}
\end{figure}

The expected phase portrait for the system is shown in fig.
\ref{PhasePort}. At constant $d/l$, as the mean filling
decreases (in real materials, these would be differently doped
samples), the ground state shifts from a SILC miniband state
through $(1,1,1)$, $(0,3,0)$,
$(1,3,1)$, $(0,5,0)$, $(1,5,1)$, and
$(0,7,0)$ liquid states; at sufficiently low $\nu_0$ a
phase transition to a staged crystalline state (shown in green)
occurs.  At physically relevant values of $t_\perp \approx 0.03 \frac{e^2}{\epsilon_0 l}$, the SILC miniband state is
the ground state only at high densities where integer-filled Laughlin
liquids would otherwise occur.

The structure of the phase portrait reflects the competition
between Hartree and Fock energy contributions.  The Fock energy
decreases with increasing density, and thus always favors maximal
staging. As $\nu_0$ decreases at fixed $d$, the Fock energy
of staging to a given filling $\nu$ remains fixed, while the
Hartree energy due to staging increases approximately linearly in
$n+1 = \nu / \nu_0$. Hence a series of transitions to states of
lower filling in the occupied layers occurs.  Once $\nu < 1/7$,
the crystalline states have lowest energy.  At fixed $\nu$, a
transition occurs between a Halperin state at higher $\nu_0$ (and
hence smaller separation $n d$ between occupied layers) and
a Laughlin liquid at lower $\nu_0$ (favored at larger $n
d$). The effect of increasing $d$ at fixed $\nu_0$ is
similar: it increases the effective cost of staging, and hence
favors less filled states; at fixed $\nu$ a phase transition
between Halperin and Laughlin liquid states occurs.

The crystalline states investigated in this paper have all assumed
`pancake' charge distributions confined to individual lattice
planes.  However, $c$-axis hopping will allow the electron charge
distribution to spread out in the $z$-direction.  For weak
hopping, the crystalline states should be stable throughout much
of their phase diagram, but eventually as $t_\perp$ increases,
states like the SILC are preferred. In the continuum, one can
define a family of variational three-dimensional crystalline
states as generalizations of the two-dimensional quantum Wigner
crystal of Maki and Zotos \cite{Maki83}, writing
\begin{equation}
\Psi={\rm det}\>\Big[\psi_{{\vec R}_i,Z_i}\big({\vec
r}_j,z_j\big)\Big]\ ,
\end{equation}
where ${\vec r}_j$ is the electron position projected onto the
$(x,y)$ plane, and $z$ is the $c$-axis coordinate, and where
\begin{equation}
\psi_{{\vec R},Z}\big({\vec r},z\big)=\phi_{\vec R}\big({\vec
r}\big)\,\varphi(z-Z)
\end{equation}
is a product of the in-plane lowest Landau level coherent state
and a trial single particle wavefunction describing the
localization of the electron along the $c$-axis.  An obvious
choice for $\varphi(z)$ would be a harmonic oscillator
wavefunction with a length scale $\lambda$ that is a variational
parameter, determined by the local curvature of the
self-consistent crystalline potential. In this paper, we {\it
underestimate\/} the stability of our crystalline states; allowing
the electrons to delocalize somewhat in the transverse dimension
will lower their energy. We will report on calculations based on
these states, and their obvious extensions to correlated Wigner
crystal states described in Eqn. \ref{psi}, in a future
publication.

We conclude that fractionally filled quantum Hall states cannot
occur as stable ground-states in isotropic materials under
realistic $\vec{B}$ fields for which the magnetic length is much
larger than the $c$-axis lattice constant unless the density is
sufficiently low so that a staged FQH state becomes energetically
favorable. At even smaller densities, the FQH staged ground-states
give way to a staged 3D Wigner crystal with large unit cells in
the direction parallel to the magnetic field.

\noindent{\it Acknowledgments\/} -- FB and BAB gratefully
acknowledge conversations with F. D. M. Haldane and S. L. Sondhi.
DPA is grateful to the support and hospitality of the Stanford
Institute for Theoretical Physics during the early stages of this
work.

\noindent\emph{Appendix:} To use the Monte Carlo method initially used by Laughlin for FQH systems \cite{laughlin1983} in a layered
system, we chose periodic boundary conditions and compute the
Coulomb energy of each configuration as a Ewald sum over repeated
copies of the fundamental cell.  Particle configurations are
generated with a probability dictated by the squared wave
function; the energy is calculated by averaging the Coulomb
potential over a large number of configurations. The fundamental
cell is a square in plane of side length $\sqrt{2 \pi l^2 N_{\rm 2D}/\nu}$,
where $N_{\rm 2D}$ is the total number of particles per
layer.  These dimensions are chosen to enclose the maximum
possible area over which the distribution of electrons in a
droplet is essentially uniform, minimizing boundary effects. At
each step in the simulation, only particles which are found within
the fundamental cell are included in calculating the energy,
though particles outside this region may return at a later Monte
Carlo step.

The vertical dimensions of the fundamental cell are determined by
the minimum number of layers which gives accurate energies. The
height of the unit cell $N_{\rm layers} d$ must be at least several times the mean in-layer inter-particle spacing; otherwise the
computed energy will be artificially high due to particles
$N_{\rm layers}+1$ layers apart which lie directly above each other.
At very small $d$ this limits the accuracy of the
simulations, as extremely large numbers of layers must be used to
obtain reasonable values.

\bibliography{QHbib}

\end{document}